\documentclass[doublecol]{epl2}

\usepackage{amsmath,amssymb}

\title{{Sublattice ordering in a dilute ensemble of
defects in graphene}}

\author{V.V. Cheainov \inst{1} \and O. Sylju\aa sen \inst{2} \and B.L. Altshuler
\inst{3} \and V.I. Fal'ko \inst{1} }
\shortauthor{V.V. Cheianov \etal}

\institute{
  \inst{1} Department of Physics, Lancaster University - Lancaster, LA1~4YB, UK \\
  \inst{2} Department of Physics, University of Oslo - PO Box 1048 Blindern, N-0316
Oslo, Norway \\
  \inst{3} Department of Physics, Columbia University -  538 West 120th Street, New York,
NY 10027, USA
}

 {Structure of graphene}

\abstract{Defects in graphene, 
such as vacancies or adsorbents attaching themselves to carbons, 
may preferentially take positions on one of its two
sublattices, thus breaking the global lattice symmetry. This leads to
opening a gap in the electronic spectrum. We show that such a sublattice
ordering may spontaneously occur in a dilute ensemble defects, 
due to the long-range interaction between them
mediated by electrons. As a result sublattice-ordered domains may form, with
electronic properties characteristic of a two-dimensional topological
insulator.}

\begin{document}

\maketitle

The flatland of graphene \cite{1} turned out to be a laboratory where
fundamental ideas of particle physics and cosmology \cite{2} find an
on-table realization. The low-energy spectrum of charge carriers in graphene
resembles that of massless relativistic particles \cite{3}: it directly
follows from the symmetry of its hexagonal lattice \cite{3,4} that the
spectrum is gapless and linear. In pristine graphene, the lattice symmetry
is so robust and resilient against spontaneous symmetry breaking that no
gap-generation mechanism has yet been proposed. Here we show that a
relatively low concentration of adsorbents sprinkled over graphene flake can
reverse the situation. It turns out that even at relatively low
concentration the adsorbents may self-organize into the partially ordered
state. The ordering generates a gap in the electronic spectrum {and
converts graphene} into a topological insulator.

The honeycomb lattice of graphene with two carbons in the unit cell can be
considered as a combination of two overlaying triangular sublattices, A and
B -- see Fig. 1. It is the symmetry between the sublattices that prevents
formation of the gap in pristine graphene \cite{3,4}. Consider now one-sided
chemisorption of atoms on a flake deposited on an insulating substrate. Some
adsorbents, such as {fluorine}, hydrogen, or hydroxyl group are known to form a
covalent bond with a particular carbon \cite{5,6,7,8}, thus violating the
symmetry between the sublattices locally. As long as the density $\rho $ of
the adsorbents is low ($\rho \ll a^{-2}$ , where $a$ is the lattice constant
of graphene) and their locations are random, one should expect approximately
equal occupation of the sublattices, $\rho _{A}\approx \rho _{B}$ (here, $%
\rho _{A/B}$ is density of adsorbents on A/B sublattice, and $\rho =\rho
_{A}+\rho _{B}$). Effect of the local symmetry violation is, then, limited
to the scattering of electrons.

However, under certain conditions the global sublattice symmetry can be
spontaneously violated ($\rho _{A}\neq \rho _{B}$), leading to the
correlations between sublattice affiliations of adatoms. These correlations
between adsorbents' positions are electron-mediated: each adsorbent creates
a sublattice-asymmetric perturbation in the electronic density (which is
equivalent to the polarization of the electron states inside the unit cell),
which in turn interacts with other adsorbents. The effective long-range pair
interaction between the adsorbents depends on whether they reside on the
same sublattice, or on different ones. This situation can be described in
terms of the two-dimensional Ising model with the direction of an Ising
`spin', $m=\pm 1$ identifying to which sublattice each given adsorbent
belongs. The finite difference $\rho _{A}-\rho _{B}=\rho M$ between the
sublattice occupations emerges, then, as an analog of the spontaneous spin
polarization $M=\left\langle m\right\rangle $ in the Ising model, and
spontaneous breaking of sublattice symmetry over mesoscopic scale areas in a
flake gives rise to the gap in the electronic spectrum, Fig.~1.

Effect of the partial ordering illustrated in Fig.~1 is determined by the
relation between the electronic de Broglie wavelength $\lambda =h /p$
and the mean distance, $\rho ^{-1/2}$ between the adsorbents. While
`high-energy' electrons with short wavelength $\lambda \ll \rho ^{-1/2}$
are insensitive to the ordering (for them the adsorbents act as individual
scatterers), for electrons with long wavelength $\lambda >\rho ^{-1/2}$
the ordering implies global symmetry-breaking field, which opens the
spectral gap, $\Delta $ [see Fig.~1]:
\begin{equation}
\varepsilon =\pm \sqrt{v^{2}p^{2}+\Delta ^{2}},\;\;\Delta =Mua^{2}\rho .
\label{spectrum}
\end{equation}
Here, $v$ is Dirac velocity of electrons, and $u$ is the electron - adatom
coupling constant. The density of states (DoS) corresponding to the gapped
spectrum Eq. (\ref{spectrum}) is
$\gamma =\gamma _{0}\theta (\varepsilon -\Delta )$ (here $\theta (z)$ is the step
function and $\gamma _{0}=2\varepsilon /\pi v^{2}\hbar ^{2}$ is DoS in
pristine graphene, which takes into account the spin and valley degeneracy).
One can say that in a unit area of the graphene flake as many as $\Delta
^{2}/(\pi v^{2}\hbar ^{2})$ states that originally resided within the gap $%
vp<\Delta $ have been transferred by the ordering to the higher energies, $%
|\varepsilon |\sim \hbar v\sqrt{\rho }$. In the absence of carriers the
conduction band is empty, while the valence band is full. The gap formation
thus reduces the sheet density of the energy by, approximately, $\Delta ^{2}%
\sqrt{\rho }/\hbar v\sim M^{2}u^{2}(a^{2}\rho )^{2}(\sqrt{\rho }/\hbar v)$
For the ordering to be thermodynamically favorable, the energy gain should
exceed the free energy cost, $k_{B}TM^{2}\rho $ related to the entropy drop
upon ordering. We therefore predict a phase transition with the sublattice
symmetry breaking at $T=T_{c}$,
\begin{equation}
T_{c}=C\frac{u^{2}}{\hbar v/a}(a^{2}\rho )^{3/2}  \label{Tc}
\end{equation}%
where $C\sim 1$ is a numerical factor which will be evaluated below. The
formation of the gap in the electronic spectrum due to the partial ordering
of adsorbents [see Fig.~1] should manifest itself in ARPES and in a
suppressed infrared light absorption.

To describe the sublattice ordering transition we use the RKKY-like approach
\cite{9} to the pair correlations between adsorbents (adatoms) and map the
problem onto the Ising model. The Hamiltonian of graphene with adsorbents
residing on carbons has the form \cite{8,11}
\begin{equation}
\hat{H}=v\mathbf{\hat{\sigma}}\cdot \mathbf{p}+ua^{2}\sum_{i}[\hat{\sigma}%
_{z}m_{i}+\hat{V}_{\mathrm{res}}]\delta (\mathbf{r}-\mathbf{r}_{i})
\label{Hamiltonian1}
\end{equation}%
The first term $\hat{H}$ describes free electrons with linear spectrum. The
second term accounts for the interaction of the electrons with the
adsorbents. It is responsible for the A-B symmetry breaking. Three Pauli
matrices, $\hat{\sigma}_{x,y,z}$ act on the sublattice components of the
electronic Bloch function. Ising `spin' $m_{i}$ determines the sublattice A (%
$m_{i}=+1$) or B ($m_{i}=-1$) that hosts given adsorbent.
The residual term $\hat V_\mathrm{res}$ contains two contributions, which
do not affect sublattice ordering. Firstly, there are two
electronic valleys in graphene, and $\hat V_\mathrm{res}$ takes into
account inter-valley scattering. Although we
dropped the valley indices in Eq. (\ref{Hamiltonian1}), everywhere below the
valley degeneracy is taken into account. Secondly, $\hat V_\mathrm{res}$
contains a "scalar" contribution describing the sublattice- and valley-
independent channel of electron scattering. Although such a contribution
does not affect sublattice ordering directly, it is helpful to notice that
it leads to a repulsive interaction between adatoms and precludes
adatom clustering discussed in \cite{Levitov}.

A remark is due here concerning the Hamiltonian \eqref{Hamiltonian1}.
Strictly speaking, the interaction term in the  \eqref{Hamiltonian1} is not well
defined due to ultraviolet problems and is written in this form for
illustrative purposes. In a more rigorous approach
$\hat \sigma_z m_i+\hat V_{\text{res}}$ should be replaced by an
energy-dependent $T$-matrix defining the long-distance asymptotic form of
the electron wave scattered off the defect at point $\mathbf r_i.$
The typical energy of electrons involved in the RKKY exchange between
impurity atoms is $\Delta\epsilon \sim \hbar v \sqrt \rho.$ If $T(\epsilon)$
is a slowly varying function of energy for $\vert\epsilon\vert < \Delta \epsilon,$
one can neglect its energy dependence altogether. Technically,
this amounts to treating the scattering terms in Eq.~\eqref{Hamiltonian1}
in the Born approximation. Under special circumstances, for instance in the presence
of resonant impurity levels at energies comparable to $\Delta \epsilon,$ such an
approximation breaks down and an alternative approach is to be used,  {\it e.g.} such as
in \cite{Levitov}.

\begin{figure}
\onefigure{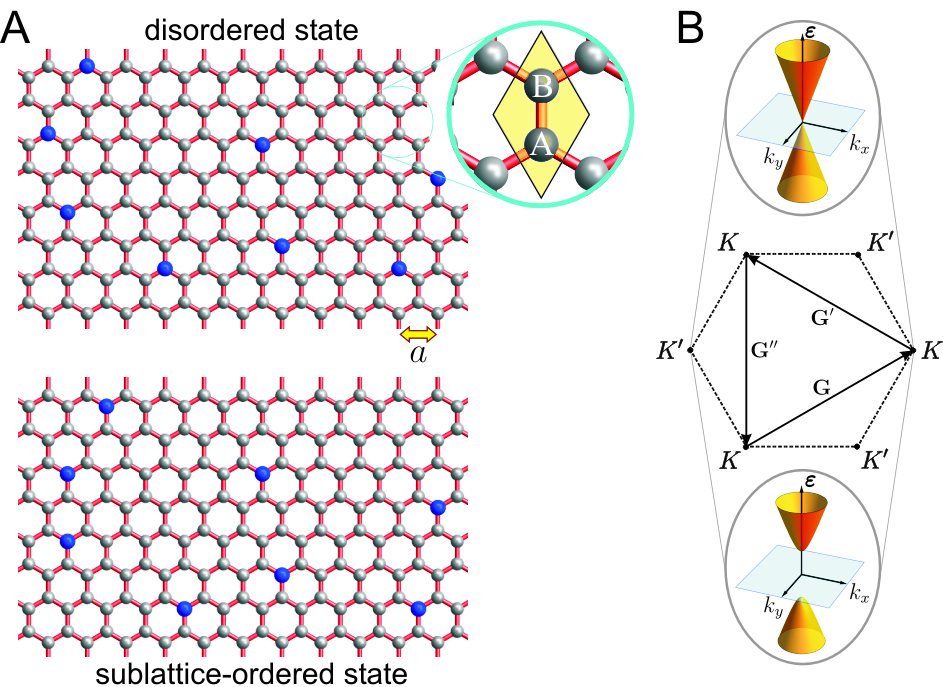}
\caption{(a) Disordered (top) and sublattice-ordered (bottom) state of a
dilute ensemble of adsorbents residing over the sites of the honeycomb
lattice of graphene. (b) The 1st Brillouin zone of graphene (dashed lines),
with three reciprocal lattice vectors linking triads of equivalent corners,
such as points $\mathbf K$. The top/bottom insets illustrate that partial
sublattice ordering opens a gap in otherwise gapless electronic spectrum of
graphene.}
\label{Fig01}
\end{figure}

An adsorbent `\textit{i}' attached to one of carbons redistributes the
electron density between A and B sites. The density redistribution creates a
sublattice-asymmetric potential landscape for another adatom `\textit{j}' at
a distance $r_{ij}$. We have found that energy of two adatoms is lower when
they reside at the same sublattice than when sublattices are different:
\bigskip
\begin{equation}
E_{\mathrm{int}}=-\frac{J}{2}\sum_{i\neq j}\frac{m_{i}m_{j}}{r_{ij}^{3}\rho
^{3/2}},\;J=\frac{u^{2}}{4\pi \hbar v/a}(a^{2}\rho )^{3/2}  \label{Eint}
\end{equation}%
and $r_{ij}=|\mathbf{r}_{i}-\mathbf{r}_{j}|$. The electron-mediated
interaction $E_{\mathrm{int}}$ between adatoms can be derived using the
standard diagrammatic technique \cite{9}. The interaction Eq. (\ref{Eint})
resembles the RKKY interaction between spins in dilute ferromagnetic
semiconductors \cite{12}. There is however an important difference specific
for undoped graphene \cite{Fertig}: coupling $E_{\mathrm{int}}$
monotonically decreases as \ $r_{ij}^{-3}$ , while RKKY coupling oscillates
and decays in two dimensions only as \ $r_{ij}^{-2}$.

Conventional RKKY interaction is caused by Friedel oscillations - the
electron density around a point defect oscillating in space with the wave
vector equal to the momentum transfer upon scattering between the extreme
points of the Fermi surface. In graphene the hexagonal Brillouin zone (BZ)
is characterized by two triads of equivalent corners, $\mathbf{K}$ and
$\mathbf{K}'$ related within each triad by the reciprocal lattice vectors,
$\mathbf G, \mathbf G',  \mathbf G^{\prime\prime}$  [see Fig. 1(b)]. In the absence
of charge carriers the Fermi line is shrank to the points $\mathbf{K}$ and\
$\mathbf K'$ which become its extremes. Perturbation of the electron Vanier
functions by an adsorbent creates oscillating charge redistribution
analogous to the Friedel oscillations. The wave numbers in this case are
equal to the reciprocal lattice vectors, e.g. $\mathbf G, \mathbf G '$, or
$\mathbf G''$. Usually the amplitude of Friedel oscillations
is proportional to the Fermi level DoS of electrons. The latter, formally,
vanishes: in neutral graphene $\gamma \propto \varepsilon =vp$, e.g., $%
\gamma (0)=0$. Nevertheless the charge density modulation still takes place.
Indeed, according to the Heisenberg uncertainty principle, given the
distance $r_{ij}$ between the two adsorbents, the characteristic electron
momentum counted from the BZ corner is $p\sim \hbar /r_{ij}$. Therefore, for
the estimation of RKKY coupling we should use $\gamma \sim (\hbar
vr_{ij})^{-1}$ . This is the source of the additional $1/r_{ij}$ factor in
the RKKY interaction between spinfull on-site impurities in graphene \cite%
{Fertig}\ and in Eq. (\ref{Eint}) describing the distance dependence of the
effective interaction between adsorbents.

Equation (\ref{Eint}) determines the Ising model in the ensemble of randomly
distributed `spins' with a long-range interaction. To evaluate the critical
temperature $T_{c}$ we simulated the transition numerically: we used 10
realizations of random Poisson distributions of $N=2\times 10^{4}$ Ising
spins on a plane interacting according to Eq. (\ref{Eint}). The
thermodynamic average, $M$ of the polarization for each realization was
computed by the cluster Monte Carlo algorithm \cite{10}. Results of this
computation presented on Fig. 2 suggest that
\begin{equation}
T_{c}\approx 13J,  \label{Tc-1}
\end{equation}%
i.e. factor $C$ in Eq. (\ref{Tc}) is $C\approx 1$.

It should be noted that, strictly speaking, Eqs. (\ref{Eint}) and
(\ref{Tc-1}) only apply in the absence of charge carriers. If doping is heavy as, e.g., in
graphene on SiC \cite{13}, it suppresses the transition. Provided that there
are more carriers than adsorbents the Fermi wave number $k_{F}$ exceeds
inverse mean distance between neighboring adsorbents $k_{F}>\sqrt{\rho }$
and conventional Friedel oscillations \cite{14}, with the wave vector $2k_{F}
$ dominate the RKKY interaction in Eq. (\ref{Eint}). Under these conditions
one should expect no ordering.

Two important remarks on the ordering transition are due here. One has to
deal with competing interactions between the adsorbents. In addition to
Ising interaction in Eq. (\ref{Eint}) between the adsorbents, each of them
causes Friedel-like oscillations of a different type: with the wave vector
equal to the distance between non-equivalent BZ corners $\mathbf{K}$ and
$\mathbf K'$. The amplitude is decaying as inverse cube of the distance from
the adsorbent. From the symmetry point of view these oscillations can be
regarded as a charge density wave superlattice with a supercell three times
as big as the unit cell. They generate an effective interaction between
adsorbents which favors AB configuration (rather than AA or BB ones) of the
closest neighbor. Such interaction is anisotropic: when projected onto one
supercell, two adsorbents determine one of the three possible directions of
A-B bonds. Due to this anisotropy for a low-density $\rho \ll a^{-2}$ random
coverage the supercell correlations are unavoidably frustrated and would not
lead to ordering. This behavior should be contrasted to that of adsorbents
which reside in the middle of honeycomb lattice hexagons (such as alkali
atoms) or adatoms residing over the C-C bonds: such adsorbents cannot
distinguish between two sublattices, but they can establish partial ordering
into a superlattice with a triple-size Kekule-type unit cell \cite{24}.
\begin{figure}
\onefigure{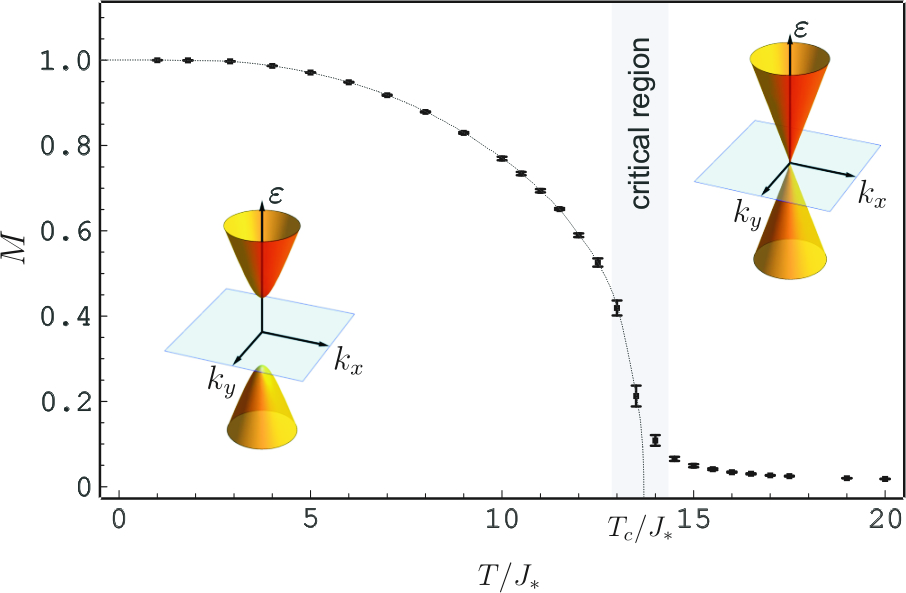}
\caption{Sublattice ordering in the ensemble of $2\times 10^{4}$ adsorbents
simulated numerically using cluster Monte Carlo algorithm \cite{10}, where
sublattice occupancies of adatoms are characterized by the `Ising spins' $
m_{i}=\pm 1$ and their thermodynamic average, $M$ is computed for 10
realizations of random Poisson distributions of spins on a square with
periodic boundary conditions interacting according to Eq. (\ref{Eint}). The
critical temperature, $T_{c}\approx 13J$, is determined approximately from
the sharp change in the `polarization' $M$. A small, but finite value of $M$
at $T>T_{c}$ is a typical finite-size effect.}
\label{Fig02}
\end{figure}

Another remark concerns the kinetics of the transition. The ordered state
arises only {if} adsorbents can hop along a flake. {This is}
possible only if the activation barrier for the adsorbent's hop between A
and B is less than the desorption barrier. Recent studies of hydrogen on
graphite \cite{15} suggest that desorption barrier for H on graphite is
lower than its diffusion barrier. Therefore hydrogenated graphene \cite{16}
may not be the likeliest candidate to observe the suggested sublattice
ordering. The alternatives are represented by {halogens, e.g. fluorine}. The best
atom/group for realization of the proposed ordering is yet to be identified.

Let us discuss electronic properties of the graphene with partially {
sublattice} ordered
adsorbents. Opening of the spectral gap offers an attractive route to
control the transport properties. It is quite likely however that the
Ising-type spontaneous symmetry breaking would result in splitting of the
flake into domains of opposite \textquotedblleft
polarization\textquotedblright\ \ $M=\pm 1$ -- regions where adsorbents
occupy a particular (A or B) sublattice. Such a domain can be viewed as a
`weak' topological insulator \cite{17,18,19}. Indeed, while inside each domain
the electron spectrum is gapped, the boundary between them supports a
mid-gap one-dimensional band of chiral electronic states \cite{20} - domain
wall states (DWS), which are similar to the quantum Hall edge states \cite%
{21}. Even in the absence of magnetic field DWS electrons in one valley,
e.g. $\mathbf K$  propagate only in one direction, whereas in the other
valley, $\mathbf K' $ the DWS on the same boundary propagate in the opposite
direction. Without intervalley scattering, the DWS would make a domain wall
an ideal ballistic `wire', with the width $w_{DW}\sim \hbar v/\Delta \sim
(\hbar v/u)(a^{2}\rho )^{-1}$ and conductance $e^{2}/h$.

Note that the sublattice-ordered adsorbents still scatter electrons between
the valleys, at a rate
\begin{equation}
\tau _{iv}^{-1}\sim \rho u^{2}a^{4}\gamma (\varepsilon )/\hbar ,\;\gamma
(\varepsilon )=2\pi /\varepsilon \hbar ^{2}v^{2}.  \label{tau-iv}
\end{equation}%
Interaction between the oppositely propagating one-dimensional DWS leads to
their localization, at the length scale of the order of the mean free path $l
$ {for intervalley scattering}.
For the energies we can substitute $\varepsilon \sim \Delta $, Eq. (\ref%
{spectrum}) into Eq. (\ref{tau-iv}) and estimate the mean free path of a DWS
as
\begin{equation*}
l\sim \frac{\hbar v}{u}\frac{(\hbar v/ua)^{2}}{(a^{2}\rho )^{2}}\gg w_{DW}.
\end{equation*}%
An immediate consequence of this would be the high resistivity of graphene
flake, with characteristic for hopping resistivity temperature dependence, $%
R\propto \frac{h}{e^{2}}\exp \{(T_{\ast }/T)^{x}\}$ ($x=\frac{1}{3}$ \ for
the Mott hoping law \cite{22}, and $x=\frac{1}{2}$ for the Efros-Shklovskii
law \cite{23}). As long as $l$ exceeds a typical domain size, this system
should also display a strong positive magnetoresistivity $R(B/B_{\ast })$
over a broad magnetic field range, $B_{\ast }\sim h/el^{2}$. The
above-described modification of transport characteristics of graphene
accompanied by opening of a gap in its optical absorption spectrum would be
natural manifestations of self-organization of a dilute ensemble of on-site
adsorbents into a sublattice-ordered state.

\acknowledgments
This study was supported by the EPSRC grant EP/G041954, US DOE contract No.
DE-AC02-06CH11357, and Notur project of the Norwegian Research Council.

\end{document}